\newcommand\rs{r_{\rm s}}
\newcommand\deltam{\delta_{\rm m}}
\newcommand\deltac{\delta_{\rm c}}
\newcommand\deltab{\delta_{\rm b}}
\newcommand\deltag{\delta_{\gamma}}
\newcommand\rseq{r_{\rm s,eq}}
\newcommand\Fg{F_{\gamma}}
\newcommand\Fb{F_{\rm b}}
\newcommand\ysc{y_{\rm sc}}
\newcommand\yeq{y_{\rm eq}}
\newcommand\Litwo{\operatorname{Li}_2}
\newcommand\omegam{\Omega_{\rm m}}
\newcommand\omegab{\Omega_{\rm b}}

\newcommand\Req{\mathcal{R}_{\rm eq}}
\newcommand\Isv{I_{\rm s, var}}
\newcommand\xlc{x_{\rm l, const}}
\newcommand\xl{x_{\rm l}}
\newcommand\xs{x_{\rm s}}
\newcommand\yd{y_{\rm d}}

\newcommand\xt{\tilde{x}}
\newcommand\deltabm{\bar{\delta}_{\rm m}}
\newcommand\deltabc{\bar{\delta}_{\rm c}}
\newcommand\deltabb{\bar{\delta}_{\rm b}}
\newcommand\deltabg{\bar{\delta}_{\gamma}}
\newcommand\ns{n_{\rm s}}
\newcommand\Tm{T_{\rm m}}
\newcommand\Gf{\mathcal{G}}
\newcommand\arcoth{\operatorname{arccoth}}

\newcommand\deltabmo{\bar{\delta}_{\rm m, out}}

\newcommand\deltabgo{\bar{\delta}_{\gamma \rm{, out}}}

\newcommand\deltabbi{\bar{\delta}_{\rm b, in}}
\newcommand\deltabgi{\bar{\delta}_{\gamma \rm{, in}}}

\newcommand\Db{D_{\rm b}}
\newcommand\Gb{G_{\rm b}}
\newcommand\Wb{W_{\rm b}}
\newcommand\hW{\hat{W}}
\newcommand\Coneb{C_{1\rm b}}
\newcommand\Ctwob{C_{2\rm b}}

\newcommand\Conepd{C_{1\rm pd}}
\newcommand\Ctwopd{C_{2\rm pd}}
\newcommand\deltamst{\bar{\delta}_{{\rm m}*}}

\newcommand\fb{f_{\rm b}}
\newcommand\Mpc{\rm\; Mpc}

\newcommand\Heq{H_{\rm eq}}
\newcommand\aeq{a_{\rm eq}}
\newcommand\cs{c_{\rm s}}
\newcommand\teq{t_{\rm eq}}

\documentclass[useAMS,usenatbib]{mn2e}

\usepackage{placeins}

\usepackage{graphicx} 
\usepackage{multirow}
\usepackage{array}

\usepackage{amsmath,amssymb}

\usepackage[T1]{fontenc}
\usepackage[latin9]{inputenc}
\usepackage{float}
\usepackage{graphicx}
\usepackage{esint}

\makeatletter
\DeclareFontEncoding{LGR}{}{}
\DeclareTextSymbol{\~}{LGR}{126}
\begin{document}

\title[Simple Linear Growth of Structure with BAO]{A Simple Analytic Treatment of Linear Growth of Structure with Baryon Acoustic Oscillations}

\author{\makeatauthor}

\author[Slepian and Eisenstein]{Zachary Slepian\thanks{E-mail: zslepian@cfa.harvard.edu} and Daniel J. Eisenstein\thanks{E-mail: deisenstein@cfa.harvard.edu}\\
Harvard-Smithsonian Center for Astrophysics, Cambridge, MA 02138\\
}
\maketitle

\begin{abstract}
In linear perturbation theory, all information about the growth of structure is contained in the Green's function, or equivalently, transfer function. These functions are generally computed using numerical codes or by phenomenological fitting formula anchored in accurate analytic results in the limits of large and small scale.  Here we present a framework for analytically solving all scales, in particular the intermediate scales relevant for the baryon acoustic oscillations (BAO).  We solve for the Green's function and transfer function using spherically-averaged overdensities and the approximation that the density of the coupled baryon-photon fluid is constant interior to the sound horizon.
\end{abstract}

\section{Introduction}
In the current consensus picture of structure formation in the Universe, Gaussian random field density perturbations created at the end of inflation grow via gravity into the large-scale structure we observe today.  On large scales, this growth is mediated by baryon acoustic oscillations (BAO) in the ionized plasma prior to decoupling $(z\sim1020)$ (Sakharov 1966; Peebles \& Yu 1970; Sunyaev \& Zel'dovich 1970; Bond \& Efstathiou 1984, 1987; Holtzmann 1989; Hu \& Sugiyama 1996; Eisenstein \& Hu 1998; Eisenstein, Seo \& White 2007 (hereafter ESW07)), Silk damping from photon diffusion (Silk 1968), and neutrino free streaming (Bond \& Szalay 1983). On smaller scales and at later times, non-linear collapse, virialization and mergers also play a role.  However, on large enough scales and at early enough times, the density fluctuations remain small relative to the background, and so the growth of structure can be accurately described using linear perturbation theory (Bernardeau et al. 2002, for a review). 

While we do not know the initial density field in any given region of the Universe, the linear-theory evolution is deterministic and is encoded in the Green's function in configuration space or the transfer function in Fourier space.  These are a Fourier transform pair: the Green's function is the response to a point-like initial overdensity in an otherwise homogeneous universe, while the transfer function is the response to a flat initial perturbation spectrum.  One can combine  the Green's function or transfer function with the statistical properties of the initial, Gaussian random density field to predict late-time observables within linear theory. In particular, linear theory predicts the 2 and 3-point correlation functions of galaxies (2PCF and 3PCF), which measure the excess probability over random of finding galaxies at a given separation or on a given triangle configuration. The Baryon Acoustic Oscillation method compares measurements of the 2PCF in different redshift slices with these predictions to constrain the Universe's expansion history, exploiting an order $1\%$ bump in the 2PCF due to the sound horizon at decoupling, $r_{\rm s}\sim 150\;{\rm Mpc}$, as a fixed scale that dilates as the Universe grows (Eisenstein, Hu \& Tegmark 1998; Seo \& Eisenstein 2003; Blake \& Glazebrook 2003; Hu \& Haiman 2003; Linder 2003; see Eisenstein et al. 2005 and Coles et al. 2005 for first detections, with previous observational hints summarized in Eisenstein et al. 1998). This method has already yielded measurements of the cosmic distance scale with $1\%$ precision with the Baryon Oscillation Spectroscopic Survey (BOSS) within the Sloan Digital Sky Survey (SDSS; Anderson et al. 2014), and is a leading lever for constraining the equation of state of dark energy (Weinberg et al. 2012) through upcoming efforts (Jain et al. 2015) such as Dark Energy Spectroscopic Instrument (DESI; Levi et al. 2013), Dark Energy Survey (DES), Euclid (Laureijs et al. 2011), Large Scale Synoptic Survey Telescope (LSST; LSST Dark Energy Science Collaboration 2012), and Wide-Field Infrared Survey Telescope (WFIRST; Spergel et al. 2013).
 
 In practice, the equations of linear perturbation theory, e.g. presented in Ma \& Bertschinger (1995), are solved using numerical codes such as CMBFAST (Seljak \& Zaldarriaga 1996) or  Code for Anisotropies in the Microwave Background (CAMB) (Lewis 2000).  These codes run in of order a few minutes on a typical personal computer and are considered to be accurate at the sub-percent level.  For a given set of cosmological parameters, they can unambiguously compute numerical tables of the linear theory predictions. However, an important area of work has nonetheless been semi-analytic fitting formulae for the transfer function, as these show in closed form what the dependence on cosmological parameters is. In particular the fitting formulae of Eisenstein \& Hu (1998; 1999) helped enable the development of the BAO method by demonstrating the sensitivity of the BAO bump to cosmological parameters.  These fitting formulae agreed with the numerical results of CMBFAST within $1\%$ at that time, though one might expect less good agreement now as details of the codes such as recombination and reionization history have changed.  These fitting formulae were derived by smoothly interpolating between analytic solutions available for the large and small-scale limits of the transfer function, corresponding to scales either much larger or much smaller than the sound horizon at matter-radiation equality.  The first work on the small scale limit was done by M\'esz\'aros (1974), and  the large-scale limit was also solved around this time, in Groth \& Peebles (1975); see also Kodama \& Sasaki (1984). Hu and Sugiyama (1996; hereafter HS96), Yamamoto, Sugiyama \& Sato (1997) and Weinberg (2002) treated the small-scale limit in greater detail, including the BAO behavior on small scales as well as the effects of neutrinos; Boyanovsky, de Vega \& Sanchez (2008) derive an analytic expression for the DM transfer function on very small scales for a variety of DM candidate particles during the matter-dominated regime.  For further discussion of previous work, see Peebles (1980), Padmanabhan (1993), Dodelson (1998), and Weinberg (2008). 
 
 However, the BAO scale, $\rs\sim 150\Mpc$, falls in the middle region where no analytic solution had been known and where the fitting formulae simply interpolate.  This middle region is also where the transfer function transitions from scale-independent growth on large scales to scale-dependent, increasingly suppressed growth on small scales due to the evolution of the radiation inhomegeneities during radiation-domination. From a practical standpoint, the availability of accurate numerical solutions means that the lack of analytic work here is not a problem either for understanding the parameter dependence of the BAO or for deriving constraints on the cosmic expansion history. However, from a theoretical and pedagogical standpoint it is disappointing that there is no analytic solution in this especially interesting  regime. Further, developing a means of solving the behavior on this scale might lead to a general approach that could be used on all scales, giving a completely analytic method of computing the full transfer function or Green's function.

In this paper, we develop an approach that allows us to compute the growth of structure on these scales, as well as on the smaller and larger scales that had previously been solved.  We offer a simple, self-consistent picture of how dark matter, baryons, and photons interplay to create the large-scale clustering of galaxies we observe today (insofar as this is accurately described by linear perturbation theory).  

In \S2, we offer a qualitative configuration-space picture that sets up what follows.  \S3 outlines the approximations used all the way through the paper and presents the expressions for the sound speed and sound horizon we require. \S4 derives the base equation for the growth of perturbations we will be solving, and calculates the solution outside the sound horizon used throughout the paper. In \S5 we solve for the growth of perturbations inside the sound horizon with massless baryons and no decoupling; \S6 incorporates massive baryons, decoupling, and photon diffusion (Silk) damping, appealing to simple numerical work. \S7 shows how \S6 may be redone perturbatively, with no appeal to numerical work, for small baryon fraction. \S8 concludes.

\section{A configuration-space picture}
Before the Universe becomes neutral at $z\sim1100$ and baryons and photons dynamically decouple ($z\sim1020$), the electrons are tightly coupled to the photons by Thomson scattering, and the protons to the electrons by Coulomb attraction.  As a simple starting point, consider the behavior of a spherically symmetric, point-like (Delta function) overdensity of dark matter, baryons, and photons set up at the origin at some very high redshift in an otherwise homogeneous universe. The Universe's response to this perturbation is the Green's function, and were the true initial density field known, convolving it with this response would provide the late-time linear theory matter distribution; for development of the Green's function picture see Bashinsky \& Bertschinger (2001; 2002).

Since the photon fluid has pressure $P_{\gamma} = \rho_{\gamma}c^2/3$, the perturbation at the origin will have greater photon pressure than its surroundings. Thus the photons (and tightly coupled baryons) will be launched outwards in a spherical pulse. The high pressure in the fluid opposes density fluctuations on scales much smaller than the sound crossing scale, giving the pulse a roughly constant density interior to the sound horizon.

\begin{figure}
\centering
\includegraphics[scale=0.35]{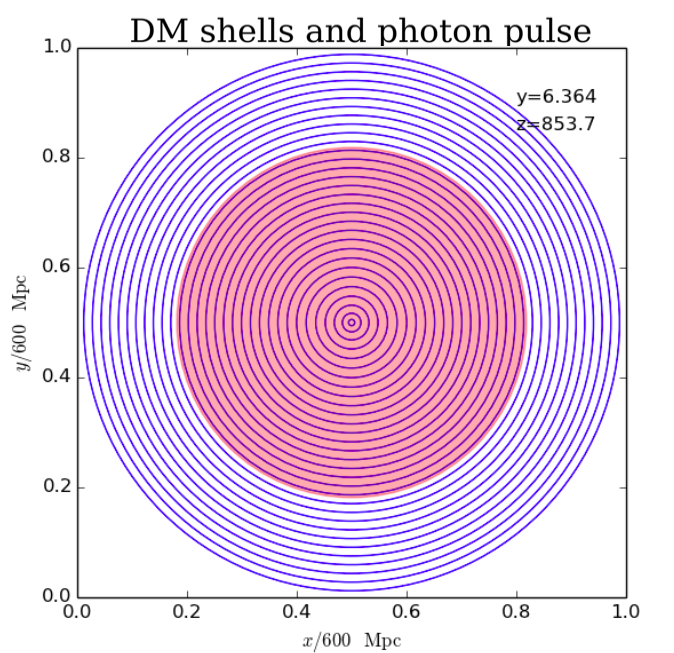}
\includegraphics[scale=0.22]{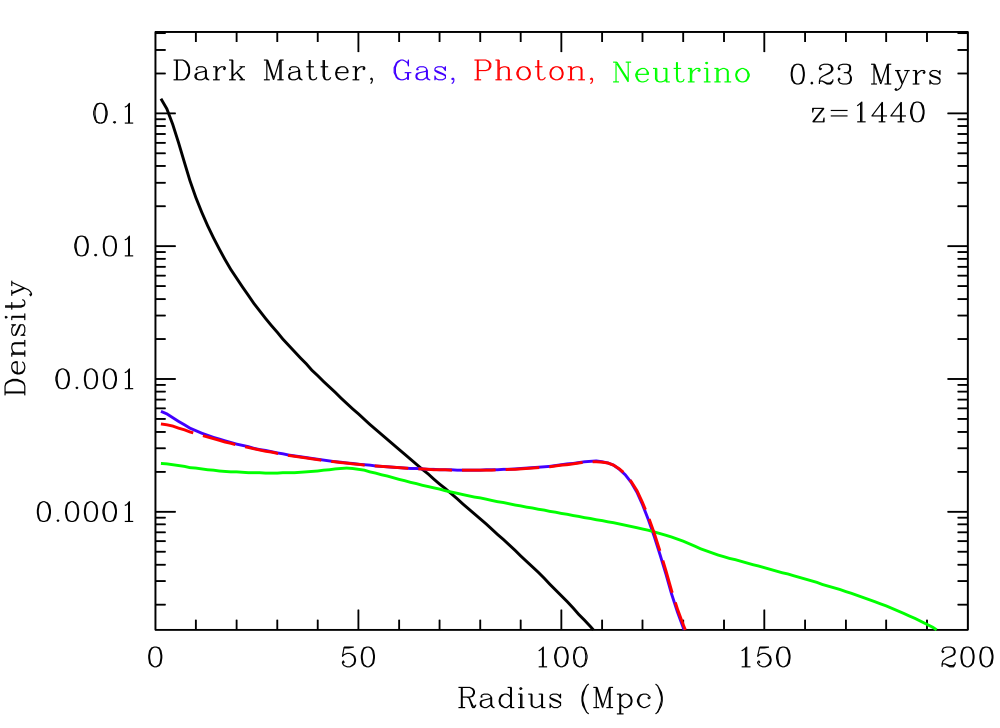}
\caption{The top panel shows DM shells in blue and the expanding photon pulse in red. For the shells outside the pulse, as regards the gravitational force it exerts the pulse may as well still be at the origin. However those DM shells the pulse has crossed receive less gravitational force inwards than their counterparts outside the sound horizon and so collapse less. The bottom panel shows the Green's functions for all the species; notice how tightly coupled the gas (baryons) and photons are, and that their spatial profile is roughly constant within the sound horizon ($\sim 130\;{\rm Mpc}$) and zero outside it. For details on the computation of this panel see ESW07.}
\label{fig:shells}
\end{figure}

We now examine the behavior of the background universe in the presence of this perturbation. First we consider a spherically symmetric bubble with radius larger than the sound horizon.  Due to the perturbation at the origin, it is overdense relative to the background universe and so will contract when measured in coordinates comoving with the background universe. All of the baryon-photon pulse is still contained within this universe, and so the average overdensity contained is the same as it was in the initial condition.  By Gauss's law, the collapse will be the same as if the baryon-photon pulse simply had remained concentrated at the origin. Effectively, this bubble does not know about the propagating pulse: the information about the pulse's behavior has only reached the sound horizon, and this bubble is sensitive to the average overdensity over a region larger than the sound horizon. Thus the collapse of this bubble is relatively straightforward to compute. Further, it is the same whether the baryons are tightly coupled to the photons or not; it depends only on the total matter density and the photon density. It is also insensitive to the details of how the sound horizon is calculated. This bubble corresponds to blue shells not yet covered by the red photon pulse in Figure \ref{fig:shells}.

Meanwhile, for a bubble of radius less than the sound horizon, mass has not been conserved: some of the baryon-photon pulse has exited the bubble, driving down the average overdensity and retarding the collapse relative to that of the outside-horizon bubble.  For a bubble that is extremely small compared to the sound horizon, the baryon-photon fluid can be treated as a homogeneous background, and this is the approximation of M\'esz\'aros (1974). However, to correctly trace the behavior of bubble whose size is similar to the sound horizon at matter-radiation equality (of order $100\Mpc$), the baryon-photon overdensity pulse's contribution to the gravitational forcing must be incorporated. How this is done certainly is sensitive to whether the baryons are counted as photons or not, as well as to details of how the sound horizon is computed. Providing an approach for following the pulse's effect is a major advance of this work. The bubbles inside the pulse correspond to those blue shells covered by the red photon pulse in Figure \ref{fig:shells}.

\section{Global approximations, definitions, and sound horizons}
Throughout this work, we make the following approximations. We ignore neutrinos entirely, and take all radiation energy density to be in photons. We assume the baryons and photons are tightly coupled, i.e. that the spatial profiles of the overdensity in each species perfectly match. Finally, we assume that the baryon-photon pulse's spatial profile is a Heaviside function in radius, constant out to the sound horizon and zero thereafter.  This is motivated by the high sound speed in the fluid, and is a good match to the exact linear theory result from CMBFAST (Figure \ref{fig:shells}, bottom panel).

We will often work in terms of the variable $y=a/\aeq$ where $\aeq$ is the scale factor at matter-radiation equality. Primes denote derivatives with respect to $y$. Where we do use time $t$, dot will denote a time derivative. $H$ is the Hubble parameter.  We will also often use the dimensionless variable $x$, where $x=r/r_{\rm s, eq}$ with $r_{\rm s, eq}$ the sound horizon at matter-radiation equality relevant within a given section (i.e. computed with either constant (\S5) or varying (\S6) sound speed. $\cs$ denotes the sound speed. Subscript $m$ will always denote total matter, subscript $b$ baryons, subscript $\gamma$ photons, and subscript $eq$ matter-radiation equality. Finally, we will use the overdensity $\delta$ and the spherically averaged overdensity $\bar{\delta}$, given respectively by
\begin{equation}
\delta(r)=\frac{\rho(r)}{\left<\rho\right>}-1,\;\;\;\bar{\delta}(r)=\frac{3}{r^3}\int_0^r{ r'^2\delta(r')dr'}.
\label{eqn:spheravg}
\end{equation}
$\rho$ is the density and $\left <\rho\right>$ is the average, background density.

In this work, we will compute the matter transfer function $\Tm(k)$ and Green's function (here $\Gf(r)$), but discuss how these relate to the power spectrum $P(k)$ or 2PCF $\xi_0(r)$ of matter. The relationships are
\begin{align}
\Tm(k) &= 4\pi \int r^2 dr j_0(kr) \Gf(r)\nonumber\\
P(k) &= Ak^{\ns}\Tm^2\nonumber\\
\xi_0(r) &=\int \frac{k^2 dk}{2\pi^2} j_0(kr) P(k),
\label{eqn:relationships}
\end{align}
where $A$ is a constant usually set by matching the observed small-scale clustering today (e.g. $\sigma_{8}$) and $\ns$ is the scalar spectral tilt, nearly $1$ (Planck Paper XIII, 2015).

In \S5 only, we treat the baryons as massless and assume that baryons and photons are fully coupled for all time. In this approximation the sound speed of the baryon-photon fluid is constant, $c/\sqrt{3}$; this is faithful to the real Universe at the $20\%$ level in $c_{\rm s}$ at $z\sim 1020$ and even better at higher redshift.  
The comoving sound horizon is then
\begin{align}
r_{\rm s,const}(y)&=\frac{c}{\sqrt{3}}\int_0^{t(y)}{\frac{dt'}{a(t')}}=\frac{c}{\sqrt{3}}\int_0^y\frac{dy'}{y'^2H(y')}\nonumber\\
&=r_{\rm s,const}(y_{\rm eq})\frac{\sqrt{1+y}-1}{\sqrt{2}-1}
\label{eqn:rs}
\end{align}
where we have normalized by the sound horizon at matter-radiation equality, $r_{\rm s, const}(y_{\rm eq})=\left[2(\sqrt{2}-1)/\sqrt{\Omega_{\rm m, 0}(1+z_{\rm eq})^3}\right](ct_{H_0}/\sqrt{3})$. The second factor corresponds to the proper sound horizon at redshift zero and the first factor rescales this to matter-radiation equality and comoving coordinates.

In \S6 and \S7 only, we take the baryons to be massive, meaning their density will dilute with the scale factor as $y^{-3}$. This also implies the sound speed varies with the ratio of  baryon to photon momentum density
\begin{align}
\mathcal{R}(y)=\frac{3}{4}\frac{\rho_{\rm b}}{\rho_{\gamma}}\approx\frac{3}{4}\frac{f_{\rm b}\rho_{\rm m}}{\rho_{\gamma}}=\Req y,
\end{align}
with $\Req=(3/4)f_{\rm b}$ and $f_{\rm b}$ the baryon fraction. Note that if we incorporated neutrinos, the first equality would remain, but the second, approximate equality would no longer hold, because the photons would no longer constitute the entire radiation energy density.

The sound speed is (Hu \& Sugiyama 1995)
\begin{align}
c_{\rm s}(y)=\frac{c}{\sqrt{3(1+\mathcal{R}(y))}},
\end{align}
and the sound horizon is (HS96)
\begin{align}
r_{\rm s,var}(y)&=c\int_0^y\frac{dy'}{y'^2H(y')\sqrt{3(1+\mathcal{R}(y)}}\nonumber\\
&=r_{\rm s,var}(y_{\rm eq})\frac{I_{\rm s, var}(y)}{I_{\rm s, var}(\yeq)}
\label{eqn:rs_var}
\end{align}
with 
\begin{align}
I_{\rm s, var}(y) = \ln\left[\frac{\sqrt{\Req+\mathcal{R}(y)}+\sqrt{1+\mathcal{R}(y)}}{1+\sqrt{\Req}} \right];
\end{align}
we have again normalized to the sound-horizon at matter-radiation equality. In \S6 decoupling of the baryons from the photons does occur and we take it to be instantaneous.

\section{Evolution outside the sound horizon}
\subsection{Perturbed shells}
We begin with the Newtonian gravitational acceleration at the surface of an overdense sphere of radius $r$ filled with matter and photons with initial densities $\rho_{\rm m,0}$ and $\rho_{\gamma,0}$:
\begin{align}
\ddot{r}& = -\frac{GM(<r)}{r^2}= -\frac{4\pi G}{3}(r_0^3 \rho_{\rm m, 0}r^{-2} + 2\rho_{\gamma,0}r_0^4r^{-3})
\label{eqn:newtons_law}
\end{align}
Multiplying by $\dot{r}$ and integrating equation (\ref{eqn:newtons_law}) with respect to $t$, we obtain 
\begin{align}
\left(\frac{\dot{r}}{r}\right)^2 =\frac{8\pi G}{3}\left( \rho_{\rm m,0}(r/r_0)^{-3} +\rho_{\gamma, 0}(r/r_0)^{-4}+C(r/r_0)^{-2}\right),
\end{align}
where $C$ is an integration constant. We are interested in the case of small differences of $\rho_{\rm m}$ and $\rho_{\gamma}$ from their background values, and we take the background Universe to be geometrically flat and contain only matter and photons (curvature and dark energy are negligible at the redshifts we consider). With this in mind, $C$ can be interpreted as the curvature produced by a perturbation to the background Universe. Consquently for small density perturbations $C$ will also be small. Setting $r_0=1$ at matter-radiation equality and defining the Hubble parameter (for the background Universe) at that epoch as 
\begin{align}
H^2_{\rm eq} = \frac{8\pi G}{3}\left( \rho_{\rm m, eq}+\rho_{\gamma,{\rm eq}}\right), 
\end{align}
equation (\ref{eqn:newtons_law}) becomes
\begin{align}
\left( \frac{\dot{r}}{r}\right)^2 = \frac{H_{\rm eq}^2}{2}\left( r^{-3}+r^{-4}+Cr^{-2}\right)
\end{align}
Taking the square root, multiplying through by $r$, and rearranging differentials, we find
\begin{align}
\int_0^{r}\frac{dr'}{\sqrt{r'^{-1}+r'^{-2}+C}}=\frac{H_{\rm eq}t}{\sqrt{2}}.
\end{align}
Multiplying the integrand's numerator and denominator by $r'$ and Taylor expanding to first order in $C$ we find
\begin{align}
\frac{H_{\rm eq}t}{\sqrt{2}}&=\int_0^r\frac{r' dr'}{\sqrt{1+r'}}- C\int_0^r \frac{r'^3dr'}{2(1+r')\sqrt{1+r'}} \nonumber\\
& = I_0(r) - CI_1(r) +\mathcal{O}(C^2)
\label{eqn:time}
\end{align}
with 
\begin{align}
I_0(r)& \equiv \frac{4}{3}+\frac{2}{3}(r-2)\sqrt{1+r} \nonumber\\
I_1(r) &\equiv \frac{16}{5}\left[\frac{1}{\sqrt{1+r}}\left(\frac{r^3}{16}-\frac{r^2}{8}+\frac{r}{2}+1\right)-1\right].
\end{align}
Now consider an overdense homogeneous bubble that is slightly perturbed from the background Universe, with  radius $r=y\left[1-\beta(y)\right]$ where $y=a/a_{\rm eq}$ is the scale factor normalized to unity at matter-radiation equality and $\beta \ll 1$. Further let $\beta(1)=\beta_{\rm eq}$. We wish to find the curvature perturbation $C$ produced by this radial perturbation. For the background Universe, as already noted there is no curvature so $C=0$, and we demand that the perturbed shell and the background Universe measure the same time of matter-radiation equality. In view of equation (\ref{eqn:time}) this means
\begin{align}
\frac{\Heq \teq}{\sqrt{2}}= I_0(1) =  I_0(1-\beta_{\rm eq}) -CI_1(1-\beta_{\rm eq}).
\end{align}
Taking Taylor series for $I_0$ and $I_1$ about unity, we find the curvature  $C$ induced by the radial perturbation $\beta_{\rm eq}$ as
\begin{align}
C =\frac{1}{I_1(1)}\frac{dI_0}{dy}\bigg|_1\beta_{\rm eq}.
\label{eqn:C_soln}
\end{align}
Indeed, the perturbed and background Universe always measure the same time, so subtracting equation (\ref{eqn:time}) for the background Universe (set $r=y$) from that for the perturbed one (set $r=y\left[1-\beta(y)\right]$), using our solution (\ref{eqn:C_soln}) for $C$,  and solving for $\beta(y)$, we find
\begin{align}
\beta(y) &= \frac{\sqrt{1+y}}{y^2}\frac{I_1(y)}{\sqrt{2}I_1(1)}\beta_{\rm eq}\nonumber\\
&=\frac{1}{23-16\sqrt{2}}\left[y-2+\frac{8}{y}+\frac{16}{y^2}-\frac{16}{y^2}\sqrt{1+y} \right]\beta_{\rm eq}.
\label{eqn:beta_out_soln}
\end{align}
This relation describes the scale-factor-dependence of the growth of an overdense region where the photons and matter move only under gravity and mass is conserved within the bubble. It is valid for all bubbles whose radii are larger than the sound horizon $\rs$.  Notice that $\lim _{y\to 0}\beta(y)=0$ and that at $y=1$ we recover $\beta(1) =\beta_{\rm eq}$. Weinberg (2002) gives a different derivation of this result for the density, and Kodama \& Sasaki (1984) find the analogous time dependence for the potential (see also Dodelson 2003).

\subsection{Spherically averaged overdensities}
Thus far we have found the time-dependence of the growth of radial perturbations outside the sound horizon. We now need to incorporate the spatial initial condition demanded by the Green's function: a Dirac-delta function perturbation at the origin, so that
\begin{align}
\deltam(r,0)&=\delta_{\rm D}^{[3]}(\vec{r})\nonumber\\
\deltag(r,0)&=\frac{4}{3}\deltam(r,0),
\end{align}
the factor $4/3$ coming from adiabaticity.
To connect these initial conditions to $\beta$ we must relate $\beta$ to $\delta$.  Given a sperical shell of radius $r_0$ with average density $\bar{\rho}_0$ equal to the background density $\left<\rho\right>$, perturbing $r_0$ to $r_0(1-\beta)$ gives a new density $\bar{\rho}=\bar{\rho}_0(V/V_0)\approx \bar{\rho}_0(1+3\beta)$. So a radial perturbation $\beta$ implies a spherically-averaged overdensity
\begin{align}
\bar{\delta}=3\beta.
\end{align}
This relation means the time-dependence of a spherically-averaged overdensity is the same as that of $\beta$. Meanwhile the initial condition on $\bar{\delta}$ is
\begin{align}
\deltabm(r,0)&=\frac{1}{4\pi r^3}\nonumber\\
\deltabg(r,0)&=\frac{4}{3}\deltabm(r,0)
\label{eqn:gen_ics}
\end{align}
using equation (\ref{eqn:spheravg}). 
Rewriting equation (\ref{eqn:newtons_law}) using $d/dt = (yH)d/dy$ and with $r_0=1$, the equation of motion for $\deltabm$ is
\begin{align}
\deltabm''+\frac{2+3y}{2y(1+y)}\deltabm' -\frac{3}{2y(1+y)}\deltabm = \frac{3}{y^2(1+y)}\deltabg;
\label{eqn:eom_outside}
\end{align}
since no spherical shells cross, adiabaticity is maintained for all time so $\deltabg$ can be replaced with  $(4/3)\deltabm$ above.

The solution is 
\begin{align}
\deltabmo(r,y) &= \deltabm(r,0)O(y)\nonumber\\
\deltabgo(r,y) &= \deltabg(r,0)O(y)
\end{align}
with 
\begin{align}
O(y) = y-2+\frac{8}{y}+\frac{16}{y^2}-\frac{16}{y^2}\sqrt{1+y}
\label{eqn:outside}
\end{align}
from equation (\ref{eqn:beta_out_soln}).

\section{Inside the sound horizon: massless baryons}
\label{sec:zerobaryons}

We now turn to bubbles with $r<r_{\rm s}$. We again begin with the Newtonian force law (recall equation (\ref{eqn:newtons_law})), but here focus on a shell of matter with radius $r_{\rm m}$:
\begin{align}
\ddot{r}_{\rm m} =-\frac{4\pi G}{3}\left(\rho_{\rm m} r_{\rm m} + 2\rho_{\gamma,{\rm in}}r_{\rm m}\right),
\label{eqn:newtons_law_inside}
\end{align}
and perturb so that $r_{\rm m}= y(1-\beta_{\rm m})$.  
Here $\rho_{\rm m}$ is simply the background matter density, but the photon density $\rho_{\gamma,{\rm in}}$ is more subtle. The photons are traveling outwards in the BAO and so some will exit our bubble: photon number {\it within the bubble} is no longer conserved.  There are three effects that alter $\rho_{\gamma,{\rm in}}$. First, the photon overdensity only (not the background density) dilutes as the sound horizon grows: the overdensity propagates to larger and larger scales and thus decreases in amplitude. Second, the dilution is somewhat balanced by the growth of the photon overdensity as its overdense bubble of radius $\rs$ contracts relative to the unperturbed background. Third, the unperturbed photon density dilutes as $y^{-4}$ as the background universe expands.

We approximate that the photon overdensity is constant within the sound horizon. This approximation is motivated by the extremely high pressure in the relativistic fluid; it quickly smoothes any inhomogeneites in the photon density within $\rs$. Further motivating this approximation, Padmanabhan (1993) shows that in a radiation-dominated cosmology the photon perturbation in Fourier space is $j_1(k\rs)/(k\rs)$, which in configuration space is a constant out to the sound horizon and zero beyond. Comparing with numerical results from CMBFAST originally presented in ESW07, and shown here in the bottom panel of Figure \ref{fig:shells}, also confirms our approximation. In detail, the photon density peaks up somewhat near the origin due to the dark matter there and also has a slight bump at the sound horizon from the convergence of the outgoing pulse with photons infalling from outside the sound horizon.  

\begin{figure}
\includegraphics[scale=.5]{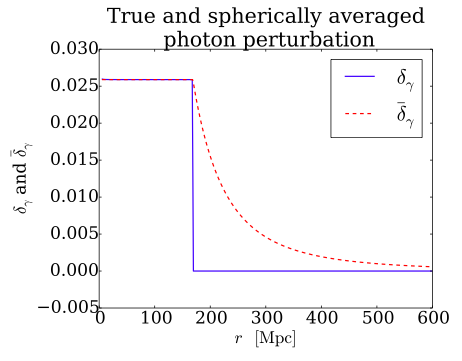}\caption{An example of passage from actual overdensity to spherically averaged
overdensity; here we show the true Silk-damped baryon perturbation (solid blue) and its spherical average (red dashed). Notice the compact support of the former but not the latter; outside the sound horizon (here evaluated at decoupling), the spherically averaged perturbation falls as $1/r^3$.}
\label{fig:spherical_avg}
\end{figure}

Since in our work the photon overdensity (and hence spherically averaged overdensity) is constant within the sound horizon, we need only set its amplitude. By construction, the spherically averaged photon overdensity must be continuous at the sound horizon, so this amplitude is simply $\deltabgo(\rs(y),y)$. Figure \ref{fig:spherical_avg} shows how the true and spherically averaged photon overdensities compare and how the latter outside the sound horizon can be used to set the amplitude within the sound horizon.

Inside the sound horizon, the dark matter perturbation's evolution is given by equation (\ref{eqn:eom_outside}): recall this simply came from Newton's law of gravity.  However, here we treat the photons as a forcing given by 
\begin{align}
\deltabgi(y)=\deltabgo(\xs(y),y)
\label{eqn:photondens}
\end{align}
rather than as following the matter as they did in \S4; we have changed variables to the dimensionless $\xs =\rs/\rseq$. We also define a forcing function 
\begin{align}
\Fg(y) = \frac{3}{y^2(1+y)}\deltabgi(y),
\label{eqn:forcing}
\end{align}
so that equation (\ref{eqn:eom_outside}) here becomes
\begin{align}
\deltabm''+\frac{2+3y}{2y(1+y)}\deltabm' -\frac{3}{2y(1+y)}\deltabm = \Fg(y).
\label{eqn:matterin}
\end{align}
We solve using variation of parameters, which writes the full solution in an eigenbasis given by solutions of the homogeneous (unforced) equation (i.e. with the righthand side of equation (\ref{eqn:matterin}) set to zero). The homogeneous equation is the M\'esz\'aros (1974) equation with a growing solution
\begin{equation}
G\left(y\right)=y+\frac{2}{3}
\label{eqn:grow}
\end{equation}
and a decaying solution (Groth \& Peebles 1975)
\begin{equation}
D\left(y\right)=\frac{3}{2}\left(3\sqrt{1+y}-\left(2+3y\right)\operatorname{arccoth}\left[\sqrt{1+y}\right]\right),
\label{eqn:decay}
\end{equation}
and the Wronskian is 
\begin{equation}
W\left(y\right)\equiv G(y)D'(y)-D(y)G'(y)=\frac{1}{y\sqrt{1+y}}.
\label{eqn:wronskian}
\end{equation}

The general solution of equation (\ref{eqn:matterin}) is then
\begin{align}
\bar{\delta}_{\rm m,in}\left(y,x\right)&=\bigg\{ G\left(y\right)\left(C_{1}\left(x\right)-\int_{y_{\rm sc}\left(x\right)}^{y}\frac{D\left(y'\right)\Fg\left(y'\right)}{W\left(y'\right)}dy'\right)\nonumber\\
&+D\left(y\right)\left[C_{2}\left(x\right)+\int_{y_{\rm sc}\left(x\right)}^{y}\frac{G\left(y'\right)\Fg\left(y'\right)}{W\left(y'\right)}dy'\right]\bigg\},
\label{eqn:vpsoln}
\end{align}
where $y_{\rm sc}\left(x\right)$ is the scale factor at which $x$ enters the sound horizon. 
As earlier noted, this writes the full solution in an eigenbasis given by the growing and decaying mode of the homogeneous equation; the integrals should simply be interpreted as projecting the forcing $\Fg$ onto this basis, with division by the Wronskian correcting for the possibility that the basis is not orthonormal ($W\to 1$ if it were).
Physically, the photon forcing drops rapidly with $y$ both since the photon density dilutes as $y^{-4}$ with the background expansion and the overdensity further falls as $\rs^{-3}(y)$ as the photon pulse expands.  Furthermore, $1/W(y)$ falls rapidly with $y$.  Thus we expect that the integrands are sharply peaked about their values at $\ysc$, when the shell of scaled radius $x$ is crossed by the sound horizon.  

We compute $\ysc$ by inverting equation (\ref{eqn:rs}), finding
\begin{equation}
y_{\rm sc}(x)=\left[(\sqrt{2}-1)x+1\right]^2-1.
\label{eqn:ysc}
\end{equation}
Both of the integrals in equation (\ref{eqn:vpsoln}) can be evaluated in closed form and are given in the Appendix (equations (. Meanwhile the constants $C_{1}$
and $C_{2}$ can be found by matching the inside and outside horizon
solutions and their first derivatives at the moment of horizon crossing.  
It should be emphasized that these constants are $x-$dependent. We write down the matching conditions below, which can be algebraically solved for $C_1$ and $C_2$. For $\bar{\delta}_{\rm m}$ itself, we have
\begin{align}
&\bar{\delta}_{\rm m,out}(y_{\rm sc}(x),x)=\bar{\delta}_{\rm m,in}(y_{\rm sc}(x),x).
\end{align}
Note that by construction (see equation (\ref{eqn:vpsoln})) the righthand side above simplifies to
\begin{equation}
\bar{\delta}_{\rm m,in}(y_{\rm sc}(x),x)=C_1(x)G(y_{\rm sc}(x))+C_2(x)D(y_{\rm sc}(x)).
\label{eqn:deltam}
\end{equation}
For the derivative $\bar{\delta}_{\rm m}'$, we have
\begin{align}
\bar{\delta}_{\rm m,out}'(y_{\rm sc}(x),x)=\bar{\delta}_{\rm m,in}'(y_{\rm sc}(x),x).
\label{eqn:deltamprime}
\end{align}
Here the righthand side simplifies to
\begin{align}
\bar{\delta}_{\rm m,in}'(y_{\rm sc}(x),x)=C_1(x)G'(y_{\rm sc}(x))+C_2(x)D'(y_{\rm sc}(x)).
\end{align}
The system (\ref{eqn:deltam}), (\ref{eqn:deltamprime}) can be solved algebraically for $C_1$ and $C_2$ as 
\begin{align}
C_1(x)&=\frac{3}{4\pi x^3}\bigg\{\frac{3 \xt^2}{(2+\xt)^2}\big(-25-4\xt(\xt+6)\nonumber\\
&+4(5+\xt(\xt+3)(\xt+4))\operatorname{arccoth}[x+1]\bigg\}
\label{eqn:C1}
\end{align}

\begin{align}
C_2(x)&=\frac{3}{4\pi x^3}\bigg\{\frac{8\xt^2}{3(\xt+2)^2}\big[5+\xt(\xt+3)(\xt+4)\big]\bigg\},
\label{eqn:C2}
\end{align}
where we have introduced the auxiliary variable $\xt \equiv (\sqrt{2}-1)x$ to simplify the form of these results.

We now obtain the late-time Green's function for the matter perturbation. At late times, only those terms in equation (\ref{eqn:vpsoln}) proportional to the growing mode are important so we may drop both $C_2$ and the second integral there.  We also take $y\to\infty$ in the upper bound of the first integral: since the integrand is sharply peaked about $\ysc$ this is a very good approximation. Both approximations are validated by comparison to exact results. Making them yields
\begin{align}
&\bar{\delta}_{\rm m,in}\left(y,x\right)\approx G\left(y\right)\left[C_1(x)-\lim_{y\to \infty}\int_{\ysc(x)}^y\frac{D(y')\Fg(y')}{W(y')} dy'\right].
\label{eqn:zerobarysoln}
\end{align}
At sufficiently late times, all scales of interest have entered the horizon, so this expression fully describes the spherically averaged matter perturbation with no need  in practice for an additional component describing super-sound-horizon shells. Finally, we must invert the spherical averaging. Differentiating equation (\ref{eqn:spheravg}) gives
\begin{equation}
\delta_{\rm m,in}\left(y,x\right)=\frac{1}{3x^{2}}\frac{d}{dx}\left[x^3\bar{\delta}_{\rm m, in}\left(y,x\right)\right].
\label{eqn:invert}
\end{equation}
Applying this prescription to equation (\ref{eqn:zerobarysoln}) and dividing out the $y$-dependence yields the matter Green's function,
\begin{align}
&\mathcal{G}(x) = \frac{\delta_{\rm m}(y,x)}{G(y)} \nonumber\\
&=\frac{3}{2\xt(2+\xt)^2}\bigg\{ -160 +\xt\bigg[ -135 - 31 \xt + 2\pi^2(2+\xt)^2\bigg]\nonumber\\
&+7\xt(2+\xt)\ln\left[ \frac{\xt}{2+\xt}\right] + \arcoth[1+\xt]\bigg[ (4 + \xt)  \nonumber\\
&\times [20 + \xt (26 + 7 \xt)] -  3 \xt (2 + \xt)^2 \bigg\{7 \ln \xt + \ln\left[ \frac{2+\xt}{256}\bigg]\bigg\}\right]\nonumber\\
&-12\xt(2+\xt)^2\Litwo\left[ \frac{\xt}{2+\xt}\right]\bigg\}.
\label{eqn:zerobaryons_greensfn}
\end{align}
The transfer function is then given by equation (\ref{eqn:relationships}); to show the results we also convert back to physical units (recall $r=xr_{\rm s,eq}$), so that $k$ has dimensions of comoving$\Mpc^{-1}$.

\begin{figure}
\includegraphics[scale=0.25]{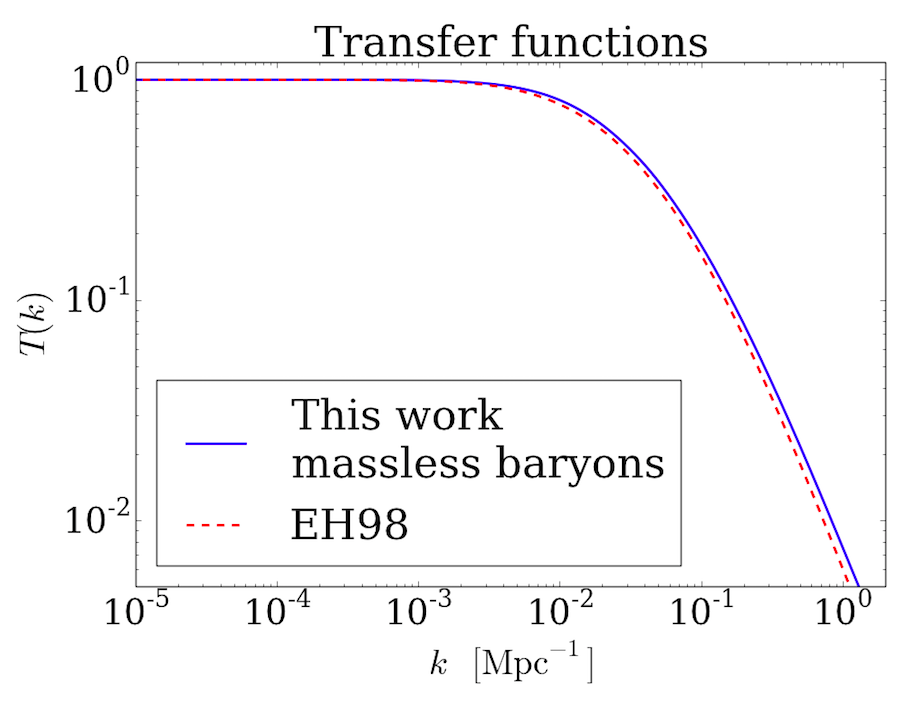}
\includegraphics[scale=0.35]{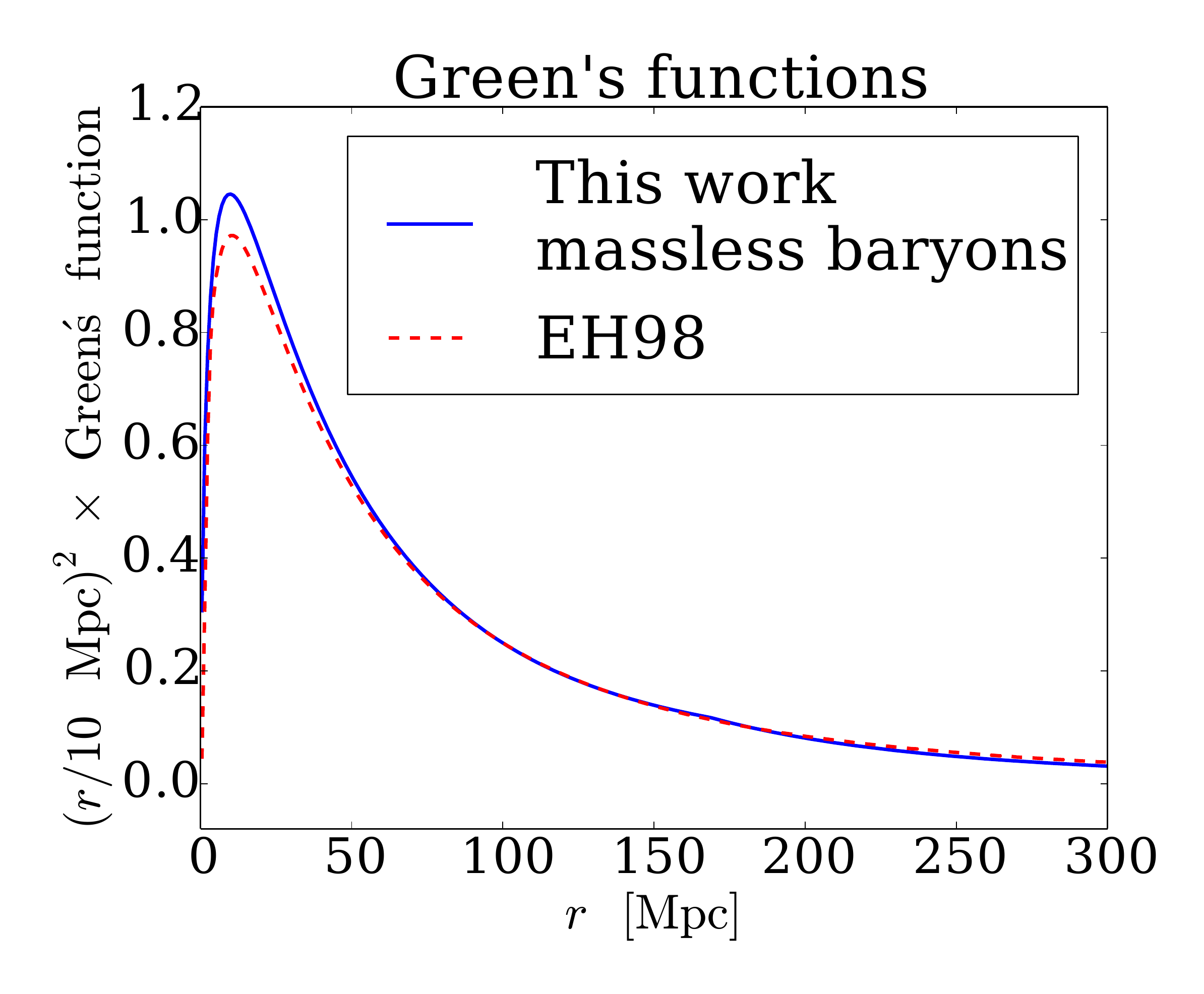}
\caption{EH98 zero baryon case is dashed red; our result is solid blue. In the top panel, the knee around $k\sim 0.1\Mpc^{-1}$ occurs since shells of scales smaller than this are crossed by the photon pulse prior to matter-radiation equality, when the photon pulse is the dominant source of the potential.  The photon pulse crosses larger shells later and later, when it is increasingly sub-dominant to the dark matter at the origin, and so these shells do not lose as much forcing as smaller shells do. Also notice that the EH98 result is below ours on small scales---this is due to its inclusion of neutrinos, which suppress small scale structure.  See the main text for further discussion of these points. The bottom panel shows the Green's function computed with both models; it is simply the inverse Fourier transform of the top panel. The slight excess of the blue curve over the red can again be attributed to additional suppression of small-scale structure in EH98 due to neutrinos.}
\label{fig:eh98}
\end{figure}

Figure \ref{fig:eh98} shows comparison in both Fourier space and configuration space of our work with EH98's ``zero baryon'' fitting formula. EH98 obtained this fitting formula by running CMBFAST (Seljak \& Zaldarriaga 1996) with constant sound speed $\cs =c/\sqrt{3}$ and trace baryon fractions and then extrapolating to zero baryons; as long as one sets the matter and photon densities consistently, this is equivalent to treating the baryons as massless as we have done here. In the transfer function, we achieve extremely good agreement with the EH98 formula for scales larger than $k\sim 0.01\Mpc$ and fairly good agreement for smaller scales.  The imperfect agreement on small scales is attributable to EH98's inclusion of neutrino free streaming, which our work ignores.   The neutrinos move at roughly $c$ outward from the density perturbation at the origin and therefore are an additional contribution
to the mass within shells larger than the sound horizon relative to the mass within shells inside the sound horizon. Since the transfer function measures the growth of perturbations relative to a shell that always remains outside the horizon ($k=0$ corresponds to a shell of infinite radius), the neutrinos additionally supress the small-scale transfer function. An area of future work might be to include neutrinos in our model. However because they free stream, their dynamics is complicated---they do not move in a coherent pulse but rather spread as they propagate into a broad ``lump'', as shown in the bottom panel of Figure \ref{fig:shells}.  The increased suppression of small-scale power in the EH98 transfer function translates to a slightly lower peak in the EH98 Green's function (Figure \ref{fig:eh98}, bottom panel).

Figure \ref{fig:eh98} reveals a knee in the transfer function around the scale of the sound horizon at matter-radiation equality. Physically, spherical shells being crossed by the sound horizon prior to matter-radiation equality genuinely notice the dilution of the radiation forcing term due to the relativistic expansion of the photon pulse. However, those being crossed by the sound horizon after matter-radiation equality do not, since by this time the radiation pulse will be dynamically sub-dominant to the DM at the origin. While this explanation for the knee in the transfer function is not new to this work (Hu \& Sugiyama 1995 originally proposed it based on the M\'esz\'aros equation), we believe articulating it in terms of the behavior of spherical shells in configuration space is better defined and more rigourous than using the language of Fourier modes ``entering the horizon'': since Fourier modes are defined throughout space, it is not actually well-defined how a mode can enter the sound horizon.  The configuration-space Green's function picture, where spherical shells really do get overtaken by the expanding photon pulse generated by the initial overdensity at the origin, we believe clarifies this idea.

\section{Inside the sound horizon: massive baryons}
\label{sec:w_baryons}
\subsection{Changing the dark matter equation of motion}
\label{subsec:changing_dm_eom}
We now take it that the baryons are massive and are released by the photons at decoupling. This imprints an additional scale on the transfer function: the sound horizon at decoupling. Here we obtain the evolution before decoupling, treating the evolution after decoupling in \S\ref{subsec:decoup}. Before decoupling, the baryons and photons are locked together. Treating the baryons as massive makes no difference to the growth of perturbations outside the sound horizon; these perturbations are only sensitive to the total mass enclosed in their shell, and so behave in the same way as in \S3.  However, for shells inside the sound horizon, treating the baryons as massive does have an effect.

Analogously to equation (\ref{eqn:photondens}) for the photon overdensity, the baryon overdensity is
\begin{equation}
\deltabbi(y)=\frac{3}{4}\deltabgi(y).
\label{eqn:bardens}
\end{equation}
Again, this is an approximation: it assumes that the baryons and photons are tightly coupled and that the pressure of the photons fully homogenizes any variation within the sound horizon. It is rather accurate, however, as the bottom panel of Figure \ref{fig:shells} shows (see also ESW07 for snapshots at several redshifts before decoupling).

Modifying equation (\ref{eqn:matterin}) by replacing $\deltabm$ with $\deltabc$ and adding baryons to the forcing, we find that the spherically averaged overdensity on shells inside the sound horizon satisfy
\begin{align}
\deltabc''+\frac{2+3y}{2y(1+y)}\deltabc' -\frac{3(1-\fb)}{2y(1+y)}\deltabc = \Fg(y) + \Fb(y)
\label{eqn:insidewbary}
\end{align}
where $\fb=\omegab/\omegam$ is the baryon fraction and we have defined the baryon forcing 
\begin{align}
\Fb(y) = \frac{3\fb y}{8}\Fg(y)
\label{eqn:baryon_forcing}
\end{align}
The driving term proportional to $\deltabm$ in equation (\ref{eqn:matterin}) came from $\rho_{\rm m}$ in equation (\ref{eqn:newtons_law}), but now only the DM density enters this driving, so we took $\rho_{\rm m} \to \rho_{\rm c}=(1-\fb)\rho_{\rm m}$, leading to the factor of $1-\fb$ on the left-hand side above.  This factor changes the homogeneous solutions from equations (\ref{eqn:grow}) and (\ref{eqn:decay}) to functions involving Gauss's hypergeometric function (see HS96 equation [D-2]). These are sufficiently complicated to prevent us from analytically projecting the forcing $\Fg+\Fb$ onto them as we did in equation (\ref{eqn:vpsoln}). Instead, we use a numerical integration up to decoupling (\S\ref{sxn:numresults}). Later, we will show how for small $f_{\rm b}\lesssim 0.5$, the case with baryons can also be solved analytically (\S\ref{sec:analytic}).
\subsection{Incorporating decoupling}
\label{subsec:decoup}
We now turn to the evolution after decoupling.  After the baryons have been released by the photons, they feel only gravity, and so are governed by precisely the same dynamics as governs the dark matter. Thus the evolution equation is now given by equation (\ref{eqn:matterin}) with no modifications.  But there is an important additional element: the initial conditions for $\bar{\delta}_{\rm m}$ are here set by integrating equation (\ref{eqn:insidewbary}) up to decoupling for $\bar{\delta}_{\rm c}$ and using equation (\ref{eqn:bardens}) for $\bar{\delta}_{\rm b}$.  The matter overdensity is the sum of the baryon and dark matter overdensities, and since the differential equation governing the evolution is linear, we may simply add them.\footnote{At decoupling there is a difference $\bar{\delta}_{\rm b}-\bar{\delta}_{\rm c}$ but one can show that it is a decaying mode.} Thus we must solve equation (\ref{eqn:matterin}) with initial conditions
\begin{align}
\bar{\delta}_{\rm m}(\yd,x) & =\fb\bar{\delta}_{\rm b}(\yd,x)+(1-\fb)\bar{\delta}_{\rm c}(\yd,x)\nonumber\\
\bar{\delta}'_{\rm m}(\yd,x) & =\fb\bar{\delta}'_{\rm b}(\yd,x)+(1-\fb)\bar{\delta}'_{\rm c}(\yd,x)
\label{eqn:ics_postdec}
\end{align}
where $\yd$ is the scale factor at decoupling.

Importantly, while the averaged baryon density is continuous, its derivative with respect to scale factor is undefined at the sound horizon, undergoing a jump from inside to outside the sound horizon.  In particular, it is
%http://tex.stackexchange.com/questions/122778/left-brace-including-several-lines-in-eqnarray
\begin{align}
\label{eqn:delta_b_deriv}
\bar{\delta}'_{\rm b}=\begin{cases}
	\frac{3}{4\pi\xs^3}\left[O'(y)-\frac{3O(y)\xs'}{\xs}\right],\;\;x<\xs;\\
	O'(y)\frac{4\pi}{3x^3},\;\;x>\xs;\nonumber\\
	{\rm undefined\; at\; }\xs\end{cases}
\end{align}
When we convert to the true density perturbation by taking a spatial derivative (see equation (\ref{eqn:invert})) we should therefore expect a Dirac-delta function like spike in the baryon density at the sound horizon.  When the baryon density evolves forward after decoupling, the growing mode inherits this spatial dependence, leading directly to the BAO feature in the late-time density. 

The continuity equation means that the time-derivative of the density connects to the velocity divergence, and so this effect---that the late-time growing mode of the baryon perturbation inherits the derivative's spatial dependence---has traditionally been known as ``velocity overshoot'' (Sunyaev \& Zel'dovich 1970; Press \& Vishniac 1980). Physically, the density's time derivative at a given radius tracks the amount of material entering or leaving that region.  The upward jump in the averaged density perturbation's derivative as we cross the sound horizon means more material is infalling towards the sound horizon at $r_{\rm s}+\epsilon$, $\epsilon$ a small positive number, than is infalling away from the sound horizon towards the origin at $r_{\rm s}-\epsilon$. Less mathematically, at a point just outside $r_{\rm s}$, outward moving (at $c_{\rm s}$) material from the baryon-photon perturbation is converging with infalling material from outside the horizon, where only gravity is important. Hence material is accumulating at the sound horizon, and when the sound horizon freezes out at decoupling, that material continues to accrete via gravity.

\subsection{Sound horizon with massive baryons}
\label{subsec:acc_sound_speed}
With massive baryons the sound horizon is given by equation (\ref{eqn:rs_var}). We now need to compute the scale factor $y_{\rm sc}(x)$ at which a given scale will enter the sound horizon; this is straightforward from inverting the dimensionless version of equation (\ref{eqn:rs_var}) for the sound horizon.  We find
\begin{align}
&y_{\rm sc}(x)=\Req^{-1}\sinh\left[ \frac{x}{\Isv(\yeq)}\right] \nonumber\\
&\times\left[2\sqrt{\Req}\cosh \left[ \frac{x}{\Isv(\yeq)}\right]  + (1+\Req)\sinh \left[ \frac{x}{\Isv(\yeq)}\right]  \right]
\end{align}
As discussed earlier, after decoupling, the photons still continue outwards. In detail the photons' behavior is now complex because they free-stream: similarly to the neutrinos they move in a broad ``lump'', illustrated in ESW07. We approximate that all of the photons are localized at a single ``light horizon'' which has propagated at $c$ subsequent to decoupling. This approximation also invokes our assumption of instantaneous decoupling.  Shells outside the sound horizon at decoupling will now enter this light horizon; it is
\begin{align}
\xl (y)=x_{\rm s}(y_{\rm d})+\left[\xlc(y)-\xlc(y_{\rm d}) \right],
\label{eq:xl}
\end{align}
with 
\begin{align}
\xlc (y)=\frac{2(\sqrt{1+y}-1)}{\Isv (\yeq)},
\label{eq:xlc}
\end{align}
where subscript $l, const$ indicates we have used a constant speed, $c$, for the photons.
The scale factor at light horizon crossing (denoted by subscript ${\rm lc}$) for a shell of radius $x$ is thus
\begin{align}
y_{\rm lc}(x)=\left(\frac{\Delta(x)x_{\rm s}(y_{\rm eq})}{2}+1\right)^2-1.
\end{align}
with $\Delta(x)\equiv x-\xs(\yd)+\xlc(\yd)$.

\subsection{Numerical results}
\label{sxn:numresults}
As earlier noted, we numerically integrate equation (\ref{eqn:insidewbary}) from our initial conditions up to decoupling. Since we want the total matter Green's function we need $\deltam(x,0) = \deltac(x,0) + \deltab(x,0)=\delta_{\rm D}^{[3]}(\vec{x})$ so our initial conditions are
\begin{align}
\deltabc(x, 0) &= \frac{3(1-\fb)}{4\pi x^3}\nonumber\\
\deltabb(x, 0) &= \frac{3\fb}{4\pi x^3}\nonumber\\
\deltabg(x,0) &= \frac{1}{\pi x^3}
\label{eqn:icswbaryons}
\end{align}

\begin{figure}
\includegraphics[scale=0.5]{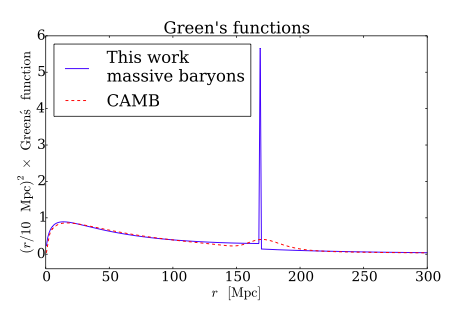}\caption{The Green's function obtained using the approach of \S \ref{sec:w_baryons}. Note the sharp spike at the sound horizon at decoupling: this is due to the discontinuity in the averaged baryon density's time derivative discussed in \S\ref{subsec:changing_dm_eom}, and is the analog of the BAO bump in the 2PCF today.  The spike has finite amplitude here because it has a minimum width set by the resolution of our spatial grid.}
\label{fig:greens_fn_nosilk}
\end{figure}

After decoupling, we numerically evolve equation (\ref{eqn:matterin}) with initial conditions given by equation (\ref{eqn:ics_postdec}) and the CDM density in that equation the solution to equation (\ref{eqn:insidewbary}).  We evolve the matter perturbation to a late enough time that even large shells have entered the photon horizon; this is our late-time Green's function.  It is shown in Figure \ref{fig:greens_fn_nosilk} compared with the result from running CAMB with {\bf no neutrinos} and the same cosmological parameters: $H_0=70$ km/s/Mpc, $\Omega_{\rm b}=0.0461224,\;\Omega_{\rm c}=0.228571,\;z_{\rm eq}=5438.48;\;T_{\rm CMB}=2.726$ K, and $f_{\rm b}=0.167905$.  The redshift of decoupling is $z_{\rm d}=1058.37$ and the sound horizon is $168.66\;{\rm Mpc}$ at that time.

Figure \ref{fig:greens_fn_nosilk} shows the results of the procedure described in \S\ref{subsec:changing_dm_eom} and \ref{subsec:acc_sound_speed}.  The plot is dominated by a sharp spike at the sound horizon at decoupling; this is due to the discontinuity in the averaged baryon density's spatial derivative discussed in \S\ref{subsec:changing_dm_eom}, and is the analog of the BAO bump in the redshift zero linear-theory 2-point correlation function.  The spike should be a delta function with zero width and infinite amplitude, but has finite amplitude by the finite width of our spatial grid so its integral remains unity.

\subsection{Adding photon diffusion (Silk) damping}
\label{subsec:silk_damping}
As Figure \ref{fig:greens_fn_nosilk} shows, our model is not in good agreement with CAMB around the sound horizon at the epoch of decoupling. This occurs because we have not incorporated an important piece of small-scale physics: photon-diffusion damping of the perturbations, also known as Silk damping (Silk 1968; see also HS96).  Silk damping will smooth the spike to a spatially extended bump at the BAO scale, and we incorporate it now using a phenomenological fitting formula from HS96, which convolves the density with a roughly Gaussian smoothing (equivalent to multiplication in Fourier space by HS96's baryon visibility function, $\mathcal{D}_{\rm b}(k)\simeq \exp[-(k/k_{\rm s})^{m_{\rm s}}]$, with $k_{\rm s}$ and $m_{\rm s}$ given by HS96 equations (E-9) and (E-10) and a 20\% phenomenological correction by EH98 equation (7)). 

The baryon visibility kernel multiplies the Fourier transform (FT) of the true baryon density. The true baryon density is the same as the spherically averaged baryon density (just a constant to the sound horizon), so its FT is
\begin{align}
{\rm FT}\{\delta_{\rm b}(y,x)\}(k)=\frac{3j_1(kx_{\rm s}(y))}{kx_{\rm s}(y))}O(y),
\label{eq:truebaryon_FT}
\end{align}
We now smooth, inverse Fourier transform, and spherically average. Denoting ``smoothed'' by subscript ``s,''
\begin{align}
&\bar{\delta}_{\rm b,s}(y,x)=\nonumber\\
&\frac{3}{x^3}\int_0^x x'^2 \int \frac{k^2 dk}{2\pi^2}\mathcal{D}_{\rm b}(k)j_0(kx')\left\{\frac{3j_1(kx_{\rm s}(y))}{kx_{\rm s}(y)} O(y)\right\}.
\end{align}

Interchanging the order of integration and evaluating the integral over $x'$ as
\begin{align}
\int_0^x x'^2 j_0(kx')dx'=\frac{x^2}{k}j_1(kx),
\end{align}
we obtain
\begin{align}
&\bar{\delta}_{\rm b,s}(y,x)=\frac{9O(y)}{xx_{\rm s}(y)}\int\frac{ dk}{2\pi^2}\mathcal{D}_{\rm b}(k)j_1(kx)j_1(kx_{\rm s}(y)).%corrected to have no k^2 in numerator; 25 march 2015.
\label{eq:silkdamp_deltab}
\end{align}

In the limit of no Silk damping, $\mathcal{D}_{\rm b}\to 1$. The $k$ integral can be done by inspection as the convolution of two boxcars evaluated at zero lag, and we recover $\bar{\delta}_{\rm b}$.

Differentiating equation (\ref{eq:silkdamp_deltab}) with respect to $y$, we find the smoothed spherically averaged baryon density's derivative as
\begin{align}
&\bar{\delta}_{\rm b,s}'(y,x)=\nonumber\\
&\frac{9O(y)}{xx_{\rm s}}\int\frac{dk}{2\pi^2}\mathcal{D}_{\rm b}(k)j_1(kx)(kx_{\rm s}')[j_0(kx_{\rm s})-\frac{2}{kx_{\rm s}}j_1(kx_{\rm s})]\nonumber\\
&+\left[\frac{9O(y)}{xx_{\rm s}} \right]'\int\frac{dk}{2\pi^2}\mathcal{D}_{\rm b}(k)j_1(kx)j_1(kx_{\rm s}).
\label{eq:silkdamp_deltabprime}
\end{align}
In the limit of no Silk damping, the second integral is again the convolution of two boxcars at zero lag, as is the term in the first integral involving $j_1(kx)j_1(k\xs)$. The term in $j_1(kx)j_0(k\xs)$ in the first integral is the convolution at zero lag of a boxcar and a Dirac delta function. Using these results recovers the no-Silk-damping limit equation (\ref{eqn:delta_b_deriv}).

\begin{figure}
\includegraphics[scale=0.5]{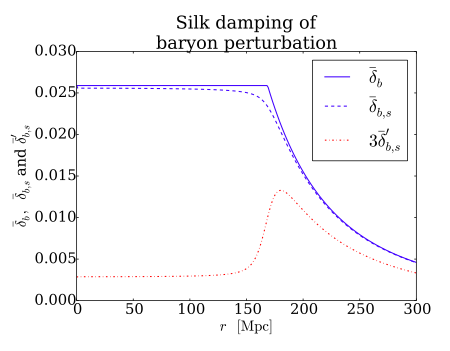}\caption{This shows the average baryon perturbation (blue solid), the Silk-damped average baryon perturbation (blue dash-dotted), and its derivative (red dash-dotted), this last multiplied by $3$ for legibility.  Notice the bump at the sound horizon in the derivative; this will lead to the BAO bump in the late-time Green's function. We do not show the no-Silk-damping derivative because it has large amplitude; it is simply a spike at the sound horizon.}
\label{fig:silk_damping}
\end{figure}

\begin{figure}
\includegraphics[scale=0.5]{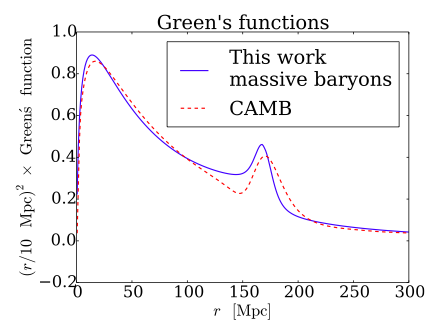}\caption{The late-time mass profile, which scales as $r^2$ times the true density perturbation. Recall the initial condition was a Dirac delta function overdensity at the origin. Note the reasonable agreement with CAMB and the BAO peak at the sound horizon at decoupling, which was $168.66\;{\rm Mpc}$ in this model; reasons for the disagreement around $150\Mpc$ are discussed in the main text.}
\label{fig:greens_silk}
\end{figure}

Figure \ref{fig:silk_damping} shows the results of applying this procedure to the spherically averaged baryon density and its derivative; as expected, we have simply smoothed the density. Note that the derivative, which before would have been a spike at the sound horizon, now simply has a smooth, relatively broad bump there. This is what sources the BAO bump in the Green's function. The Green's function with Silk damping is shown in Figure \ref{fig:greens_silk}; notice by comparing with Figure \ref{fig:greens_fn_nosilk} that the spike at the sound horizon has been smoothed into a wider bump. We find reasonably good agreement with CAMB except slightly within the sound horizon, where the difference briefly becomes of order $30\%$. We believe this is due to our approximation of instantaneous decoupling combined with mass conservation. The former means that, in our approach, when decoupling occurs, the baryons simply halt. In reality, the sound speed does not drop precipitously to zero but does so more gradually, meaning the baryons can continue to propagate slightly beyond the sound horizon at decoupling. This in turn reduces the baryonic mass within the sound horizon. Thus one would expect CAMB's Green's function to exceed ours just outside the sound horizon and be below ours just within it. It is also likely that CAMB's implementation of Silk damping has evolved since the HS96 (with EH98 corrections) fitting formula we used for the baryon visibility.  Note that neutrinos cannot be the cause of this disagreement as we ran CAMB with zero neutrinos.

\begin{figure}
\includegraphics[scale=0.5]{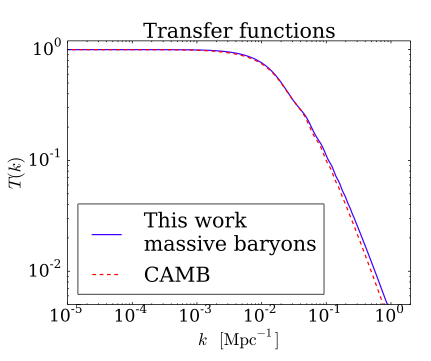}\caption{The late-time transfer function. We see reasonable agreement with CAMB, but in detail our model does not quite follow the small wiggles around the knee in the transfer function, as one might expect given the disagreement in Figure \ref{fig:greens_silk}. There is also insufficient suppression of small scale power in our model; these points are further discussed in the main text.}
\label{fig:transf_fn_full}
\end{figure}

Figure \ref{fig:transf_fn_full} shows the transfer function we obtain using the same procedure as in equations (\ref{eqn:invert}) and (\ref{eqn:relationships}). Note the rough agreement with CAMB's results. The fact that this Figure looks to show greater agreement than Figure \ref{fig:greens_silk} emphasizes the dangers of looking only in Fourier space or only in configuration space. In detail our model is not quite following all of the wiggles around $k\gtrsim 0.1\Mpc^{-1}$; this corresponds to the disagreement in Figure \ref{fig:greens_silk}.  CAMB's result also has slightly greater suppression of small-scale modes than does ours (e.g. $k>0.5\Mpc^{-1}$; this might suggest the Silk damping in our model slightly underestimates the true effect. Better understanding the physical causes of this disagreement might be a useful direction of future work.

\section{Perturbative analytical approach for small baryon fraction}
\label{sec:analytic}
Prior to decoupling, for the massive baryon-case one has the formal variation of parameters solution for the inside-horizon dark matter shells as
\begin{align}
\bar{\delta}_{\rm c,in}\left(y,x\right)&=\bigg\{ \Gb \left(y\right)\left(C_{\rm 1b}\left(x\right)-\int_{y_{\rm sc}\left(x\right)}^{y}\frac{\Db \left(y'\right)\left[\Fg \left(y'\right) + \Fb\left(y'\right)\right]}{\Wb \left(y'\right)}dy'\right)\nonumber\\
&+\Db \left(y\right)\left[C_{\rm 2b}\left(x\right)+\int_{y_{\rm sc}\left(x\right)}^{y}\frac{\Gb \left(y'\right)\left[\Fg\left(y'\right) + \Fb\left(y'\right)\right]}{\Wb \left(y'\right)}dy'\right]\bigg\},
\label{eqn:vpsoln_wbary}
\end{align}
where $\Gb$ and $\Db$ are the general solutions of the homogeneous equation (equation (\ref{eqn:insidewbary}) with the right-hand side set to zero). They are (HS96)
\begin{align}
&\Gb (y)=(1+y)^{-\alpha_1}F(\alpha_1,\alpha_1+\frac{1}{2},2\alpha_1+\frac{1}{2};\frac{1}{1+y})\nonumber\\
&\Db (y)=-\frac{2}{5}(1+y)^{-\alpha_2}F(\alpha_2,\alpha_2+\frac{1}{2},2\alpha_2+\frac{1}{2};\frac{1}{1+y}).
\label{eqn:hypergeom}
\end{align}
$F$ is Gauss's hypergeometric function, also sometimes denoted $_2 F_1$. Here the decaying solution is multiplied by $-2/5$ relative to that of HS96 (equation [D-4]) so that the limit as $\fb \to0$ agrees with our equation (\ref{eqn:decay}).
Following HS96 we have defined
\begin{align}
\alpha_i=\frac{1\pm\sqrt{1+24(1-f_{\rm b})}}{4},
\end{align}
with $-,+$ for $i=1,2$.

\begin{figure}
\includegraphics[scale=0.5]{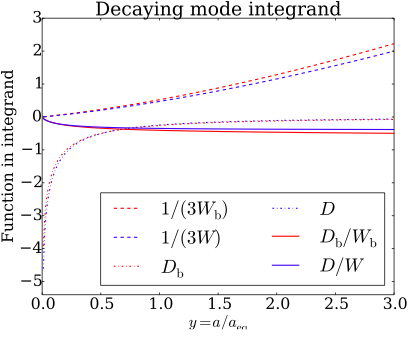}\caption{The integrand involving the decaying mode, $\Db/\Wb$ ($D/W$ with no baryons) and the various functions that enter it (we have divided $1/W$ and $1/\Wb$ by 3 for legibility). The agreement between the massless-baryon case and the case with baryons is better during radiation domination, when the integral is most important. By the time we enter matter-domination and the approximation is less good, the effects of the baryon-photon pulse will be dominated by the dark matter's gravity. This integral enters the solution multiplied by the growing mode, so it is the dominant term at late times.}
\label{fig:decay_int}
\end{figure}

The complicated form of $\Gb$ and $\Db$ means we cannot do the integrals of equation (\ref{eqn:vpsoln_wbary}) analytically as we did in the massless-baryon case (equation (\ref{eqn:vpsoln})). However, for $\fb \lesssim 0.5$, the simpler homogeneous equation solutions (\ref{eqn:grow}, \ref{eqn:decay}) for the massless baryon case (\S\ref{sec:zerobaryons}) well-approximate equations (\ref{eqn:hypergeom}).  In particular, the decaying solution above $\Db$ is very well-approximated by the massless-baryon result $D$.  This is important because the integral over $D$ is the one that gets multiplied by the growing mode in equation (\ref{eqn:vpsoln}), so its spatial dependence dominates the late-time perturbation.  Furthermore, the integrand is sharply peaked about the smallest $y$ value that enters, and for a given range in $y$ this is where the fractional error of replacing $\Db$ with $D$ is minimized.  These points are shown in Figure \ref{fig:decay_int}.  The Figure also indicates that where the difference between the Wronskians $W$ and $\Wb$ grows the decaying modes die off, controlling any error due to replacing $\Wb$ with $W$. Further, as already noted, the forcing function also dies off as $y$ grows, meaning differences between $\Db/\Wb$ and $D/W$ will enter the integral less strongly as $y$ grows.

\begin{figure}
\includegraphics[scale=0.5]{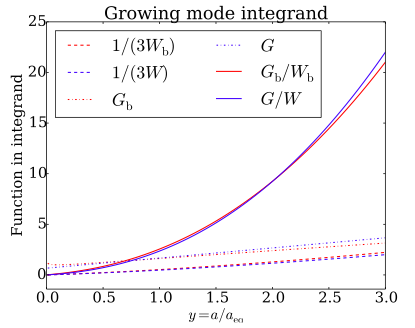}\caption{The integrand involving the growing mode, $\Gb/\Wb$ ($G/W$ with no baryons) and the various functions that enter it (we have divided $1/W$ and $1/W_{\rm b}$ by 3 for legibility). The agreement between the massless-baryon case and the case with baryons is better during radiation domination, when the integral is most important. Further, this integral enters the variation of parameters solution mutliplied by the decaying mode, so at late times its contribution is strongly suppressed. Nonetheless it is required to determine the matching constants at decoupling.}
\label{fig:grow_int}
\end{figure}

The behavior of the fractional error in the integral over $G$ is more complicated but it is always $\lesssim 20\%$ and as we have argued this integral will not strongly enter the late-time evolution. Its behavior is shown in Figure \ref{fig:grow_int}. The difference between the integrand $\Gb/\Wb$ and $G/W$ will be suppressed at late times both by the decaying mode multiplying the integral and by the rapid decrease in the forcing function that also enters the integral.

%Approximations for decaying-mode integrand and growing-mode integrand.

With these points in mind we approximate $\Gb\to G$ and $\Db \to D$ {\bf in the integrands only} in equation (\ref{eqn:vpsoln_wbary}). We can then perform the integrals. Equation (\ref{eqn:I_grow_zero_baryons}) of the Appendix  gives the integral over $G(y')\Fg(y')$ and equation (\ref{eqn:I_grow_baryons_only}) the additional integral over $G(y')\Fb(y')$; that over $D(y')\Fg(y')$ is equation (\ref{eqn:I_decay_zero_baryons}) and that over $D(y')\Fb(y')$ equation (\ref{eqn:I_decay_baryons_only}).

We can then solve for $C_{1{\rm b}}$ and $C_{2{\rm b}}$ algebraically from the matching conditions
\begin{align}
(1-f_{\rm b})\bar{\delta}_{\rm m,out}=\bar{\delta}_{\rm c,in}
\end{align}
where the righthand side again simplifies as in equation (\ref{eqn:deltam}) but now with $G\to G_{\rm b}$ and $D\to D_{\rm b}$.
For the derivative $\bar{\delta}_{\rm c}$ we have
\begin{align}
(1-f_{\rm b})\bar{\delta}'_{\rm m,out}=\bar{\delta}'_{\rm c,in}(y_{\rm sc}(x),x),
\end{align}
analogous to equation (\ref{eqn:deltamprime}).  However here the simplification for the righthand side we used after (\ref{eqn:deltamprime}) is no longer valid and we find
\begin{align}
&\bar{\delta}'_{\rm c,in}(y_{\rm sc}(x),x)=\bigg[-\frac{G_{\rm b}(y)D(y)[\Fg(y)+\Fb(y)]}{W(y)}\nonumber\\
&+\frac{D_{\rm b}(y)G(y)[\Fg(y)+\Fb(y)]}{W(y)}\nonumber\\
&+G_{\rm b}'(y)C_{1{\rm b}}(x)+D_{\rm b}'(y)C_{2{\rm b}}(x)\bigg]\bigg|_{y=y_{\rm sc}(x)}.
\end{align}
Previously the analogs of the first two terms on the righthand side canceled off.  Nonetheless the whole motive for our approximation of the integrands using the simpler growing and decaying modes was that these approximate integrands do not differ much from the true integrands; and it is just making this approximation again to cancel off the first two terms above. $\Coneb$ and $\Ctwob$ can be written compactly as
\begin{align}
&\Coneb(x)=\frac{\hW(\Db(y),(1-\fb)\bar{\delta}_{\rm m,out}(y))}{\hW(\Gb(y),\Db(y))}\bigg|_{y=y_{\rm sc}(x)}\nonumber\\
&\Ctwob (x)=\frac{\hW(\Gb(y),(1-\fb)\bar{\delta}_{\rm m,out}(y))}{\hW(\Gb(y),\Db(y))}\bigg|_{y=y_{\rm sc}(x)}
\label{eqn:vpconstswbary}
\end{align}
where $\hW$ takes the Wronskian with respect to $y$ of the two functions in its argument. 

These describe the DM perturbation after sound horizon entry but before decoupling, and give the initial condition on the sub-sound-horizon DM to match onto the total matter's evolution post-decoupling. There will still be some DM and baryons outside the sound horizon at decoupling, but these are given exactly by the outside horizon solution of \S\ref{sec:zerobaryons}. 

 After decoupling the total matter perturbation's evolution is according to equation (\ref{eqn:matterin}) with solution given by equation (\ref{eqn:vpsoln}) but where the constants, now denoted $\Conepd$ and $\Ctwopd$, are set by matching via equation (\ref{eqn:ics_postdec}). We find
\begin{align}
&\Conepd (x)=\frac{\hW(D(y),\deltamst(y))}{\hW(G(y),D(y))}\bigg|_{y=y_{\rm sc}(x)}\nonumber\\
&\Ctwopd (x)=\frac{\hW(G(y),\deltamst(y))}{\hW(G(y),D(y))}\bigg|_{y=y_{\rm sc}(x)}
\label{eqn:vpconstpostdec}
\end{align}
$\deltamst$ is the total matter perturbation before decoupling:
\begin{align}
\deltamst (y,x)=(1-\fb)\bar{\delta}_{\rm c}(y,x)+\fb \bar{\delta}_{\rm b}(y,x),
\end{align}
with $\bar{\delta}_{\rm c}$ given by equations (\ref{eqn:vpsoln_wbary}) and (\ref{eqn:vpconstswbary}) and $\bar{\delta}_{\rm b}$ by equation (\ref{eqn:bardens}).
To compute the total matter transfer function we make the same approximations as in \S\ref{sec:zerobaryons} to find
\begin{align}
\bar{\delta}_{\rm m}(y,x) &\approx\nonumber\\
& \Gb (y)\bigg[\Conepd(x)-\int_{y_{\rm sc}(x)}^\infty\frac{G(y')[\Fg(y')+\Fb(y')]}{W(y')}dy'\bigg].
\end{align}
We do not explicitly compute $\Conepd$ or invert the spherical averaging as the expressions involved would be lengthy, but it would be straightforward to do so analytically. Unfortunately we suspect that the integral against $j_0(kr)$ required to this result into a transfer function via equation (\ref{eqn:relationships}) will not be available in closed form, since it was not even for the simpler, massless-baryon case.

\section{Conclusion}
In this paper, we have developed an approximate, largely analytic approach to deriving the matter transfer function on all scales.  For a massless-baryon cosmology, we obtained a closed form for the matter Green's function and good agreement with EH98's ``zero'' or massless-baryon fitting formula for the transfer function. For the more realistic case of massive baryons, we find reasonable agreement with CAMB's results in the absence of neutrinos. For small baryon fraction, we also offer a fully analytic, approximate approach for that would with some algebra give a closed form for the matter Green's function. Up until now, there has been no analytic framework for resolving the behavior of shells being overtaken by the sound horizon around matter-radiation equality (e.g. Dodelson (2003) and Weinberg (2008) for this point). Our approach fills this gap: it is a  self-contained, reasonably realistic means of solving the linear regime of structure formation on all scales.

Our approach exploits the fact that the spherically averaged matter overdensity outside the sound horizon can be easily computed because the only force it feels is gravity.  All species (except neutrinos, which we ignore in this work) behave the same outside the sound horizon, so the spherically averaged matter overdensity also gives the photon and baryon overdensities in this region. Motivated by the high pressure in the baryon-photon fluid and working in the tight-coupling approximation, we model the baryon and photon overdensities inside the sound horizon as constant. Since the spherically averaged density perturbations are continuous, the outside horizon solution can be used to set this constant. From this we derive a closed form for the massless-baryon-case Green's function, and two ordinary linear differential equations for, respectively, the pre-decoupling evolution of the DM and the post-decoupling evolution of the total matter. 

In our model, during a given epoch there are only two components: before decoupling, the baryon-photon fluid and the dark matter, and after decoupling, the photons and the total matter. Numerically integrating these equations offers all the qualitative features of the CAMB Green's function: small scale clustering that matches and a peak at $\rs$.  Our simple implementation of a phenomenological Silk damping prescription improves the match with CAMB results.  

Finally, we have shown that for small baryon fraction $\fb$, the growing and decaying modes in the presence of baryons can be approximated by the growing and decaying modes for the massless-baryon case.  This allows approximate closed-form solution of the differential equation governing the matter overdensity in the presence of baryons.  Algebraic manipulations of these results would lead to a closed form for the matter Green's function, though these expressions would likely be cumbersome and so are not presented here.

Two major physical themes are illustrated by our work.  The first is that, even in the absence of baryons, the transfer function develops a smooth knee at the scale of the sound horizon at matter-radiation equality.  Shells outside the horizon at this epoch grow more than shells that have entered the horizon, because these latter are losing some of the gravitational force of the photon perturbation since it is partially outside their Gaussian spheres.  As the Universe expands and the photon density becomes sub-dominant to the matter density, the exact moment at which a mode enters the horizon becomes unimportant so the transfer function flattens out. In other words, after matter-radiation equality, shells entering the horizon still lose some of the photon perturbation's gravitational forcing, but this perturbation is so diluted that they barely notice.

The second physical theme relates to the BAO.  In this work we have shown that the discontinuity in the time-derivative of the spherically averaged baryon overdensity at the sound horizon at decoupling leads to a Dirac delta function-like spike in the spatial behavior of the true baryon overdensity at decoupling.  This in turn contributes to the total matter overdensity proportional to the growing mode, so the late-time transfer function and Green's function retain structure at the sound horizon.  This structure is particularly striking in the Green's function, manifesting as a localized BAO bump. This is the analog of the BAO bump seen in the matter correlation function today.  

Neither of these physical themes are new to this work (see for instance Sunyaev \& Zel'dovich 1970; Press \& Vishniac 1980; Hu \& Sugiyama 1995; HS96; EH98; Weinberg 2002; ESW2007, and many others) but we believe the configuration space, spherical shell approach presented here significantly clarifies what is meant by Fourier modes entering the horizon. Further, we believe the self-consistent, complete, and nearly fully analytic treatment offered here of the growth of structure on all scales is novel and of intrinsic interest. While we do not suggest that the work here is accurate enough to supercede either codes such as CAMB and CMBFAST or fitting formula such as EH98, we believe it is complementary. The codes are complicated and the fitting formulae are purely phenomenological in the intermediate region where they interpolate: our treatment here involves only a few equations to show how the underlying physics propagates through to the final functional forms. 

A future direction of our own work may be to extend the treatment here to neutrinos. As earlier noted, their dynamics are complicated because they free-stream. Given that upcoming large-scale structure surveys promise to place tight constraints on the sum of the neutrino masses (Jain et al. 2015), it may be rewarding to build the qualitative yet detailed understanding of neutrinos' effects our approach here might afford.

\section*{Acknowledgments}
ZS thanks Aaron Bray, Anthony Challinor, George Efstathiou, Douglas Finkbeiner, Lehman Garrison, Margaret Geller, Avi Loeb, Ramesh Narayan, Stephen Portillo, David Spergel, and Alexander Wiegand for many valuable discussions, and Meredith MacGregor, Philip Mocz, and Stephen Portillo for careful reads of the manuscript. ZS especially thanks Yuan-Sen Ting for an extensive and rigorous set of comments and Harry Desmond for helpful suggestions on the broader context of the work. This material is based upon work supported by the National Science Foundation Graduate Research Fellowship under Grant No. DGE-1144152; DJE is supported by grant DE-SC0013718 from the U.S. Department of Energy.

\section*{References}

\noindent Bashinsky S \& Bertschinger E, 2001, PRL 87, 8. %arXiv:astro-ph/0012153

\noindent Bashinsky S \& Bertschinger E, 2002, PRD 65, 12.%, arXiv:astro-ph/0202215v2.

\hangindent=1.5em
\hangafter=1
\noindent Bernardeau F, Colombi S, Gazta\~{n}aga E \& Scoccimarro R, 2002, Phys. Rep., 367, 1.

\hangindent=1.5em
\hangafter=1
\noindent Blake C \& Glazebrook K, 2003, ApJ 594, 2, 665-673.%http://adsabs.harvard.edu/abs/2003ApJ...594..665B

\noindent Bond JR \& Efstathiou G, 1984, ApJ 285, L45.

\noindent Bond JR \& Efstathiou G, 1987, MNRAS 226, 655-687.

\hangindent=1.5em
\hangafter=1
\noindent Bond JR \& Szalay AS, 1983, ApJ 274, 443-468.%http://adsabs.harvard.edu/abs/1983ApJ...274..443B

\hangindent=1.5em
\hangafter=1
\noindent Boyanovsky D, de Vega HJ \& Sanchez NG, 2008, PRD 78, 6, 063546.%http://adsabs.harvard.edu/abs/2008PhRvD..78f3546B

\hangindent=1.5em
\hangafter=1
\noindent Cole S et al., 2005. MNRAS 362, 2, 505-534.%http://adsabs.harvard.edu/cgi-bin/bib_query?arXiv:astro-ph/0501174

\hangindent=1.5em
\hangafter=1
\noindent Dodelson S, 2003, Modern Cosmology, Academic Press: Amsterdam%equation 7.25 and on recap KS1984 solution for outside-horizon potential.

\hangindent=1.5em
\hangafter=1
\noindent Eisenstein DJ, Hu W, Silk J \& Szalay AS, 1998, ApJ 494:L1-L4.%http://background.uchicago.edu/~whu/Papers/bump.pdf

\hangindent=1.5em
\hangafter=1
\noindent Eisenstein DJ \& Hu W, 1998, ApJ 496, 605. %Baryonic Features in the Matter Transfer Function.

\iffalse
\hangindent=1.5em
\hangafter=1
\noindent Eisenstein DJ, Hu W \& Tegmark M, 1998, ApJ 518, 1, 2-23.%http://adsabs.harvard.edu/cgi-bin/bib_query?arXiv:astro-ph/9807130
\fi

\hangindent=1.5em
\hangafter=1
\noindent Eisenstein DJ, Hu W \& Tegmark M, 1998, ApJ  504:L57-L60.%http://background.uchicago.edu/~whu/Papers/hubble.pdf

\noindent Eisenstein DJ et al., 2005, ApJ 633:560-574.% arXiv:astro-ph/0501171.

\noindent Eisenstein DJ \& Hu W, 1997, ApJ 511, 5.%astro-ph/9710252v1.pdf

\hangindent=1.5em
\hangafter=1
\noindent Eisenstein DJ, Seo H-J \& White M, 2007, ApJ  664:660-674.

\hangindent=1.5em
\hangafter=1
\noindent Groth E \& Peebles PJE, 1975. A \& A 41, 143-145.%http://articles.adsabs.harvard.edu/cgi-bin/nph-iarticle_query?1975A%26A....41..143G&defaultprint=YES&filetype=.pdf

\hangindent=1.5em
\hangafter=1
\noindent Holtzmann JA, 1989, ApJS 71,1.

\hangindent=1.5em
\hangafter=1
\noindent Hu W \& Haiman Z, 2003, PRD 68, 6, 063004.%http://adsabs.harvard.edu/abs/2003PhRvD..68f3004H

\noindent Hu W \& Sugiyama N, 1995, ApJ 444:489-506.%http://background.uchicago.edu/~whu/Papers/hs95a.pdf

\noindent Hu W \& Sugiyama N, 1996, ApJ 471:542-570.% arXiv:astro-ph/9510117.

\noindent Jain B. et al., 2015, preprint (arXiv:1501.07897v2)

\hangindent=1.5em
\hangafter=1
\noindent Kodama H \& Sasaki M, 1984, Prog. Theor. Phys. Supplement 78 1-166.%http://ptps.oxfordjournals.org/content/78/1.full.pdf+html

\noindent Lewis A, 2000, ApJ, 538, 473.

\hangindent=1.5em
\hangafter=1
\noindent Linder, EV, 2003, PRD 68, 8, 083504.%http://adsabs.harvard.edu/abs/2003PhRvD..68h3504L

\hangindent=1.5em
\hangafter=1
\noindent Ma C-P \& Bertschinger E, 1995, ApJ 455, 7.%http://adsabs.harvard.edu/cgi-bin/bib_query?arXiv:astro-ph/9506072

\hangindent=1.5em
\hangafter=1
\noindent M\'esz\'aros P, 1974. A \& A 37, 225-228.%http://articles.adsabs.harvard.edu/cgi-bin/nph-iarticle_query?1974A%26A....37..225M&data_type=PDF_HIGH&whole_paper=YES&type=PRINTER&filetype=.pdf

\iffalse
\hangindent=1.5em
\hangafter=1
\noindent Mukhanov V, 2004, Int. J. Theor. Phys. 43, 3, 623-668.%http://adsabs.harvard.edu/cgi-bin/bib_query?arXiv:astro-ph/0303072
\fi

\iffalse
\hangindent=1.5em
\hangafter=1
\noindent Nusser A, 2000, MNRAS 317, 902-906.
\fi

\hangindent=1.5em
\hangafter=1
\noindent Padmanabhan T, 1993, Structure Formation in the Universe. University Press: Cambridge, UK.

\noindent Peebles PJE \& Yu JT, 1970, ApJ 162, 815.

\hangindent=1.5em
\hangafter=1
\noindent Peebles PJE, 1980, Large-Scale Structure of the Universe. University Press: Princeton, NJ.

\noindent Planck collaboration, Paper XIII, 2015, arXiv:1502.01589.%http://adsabs.harvard.edu/cgi-bin/bib_query?arXiv:1502.01589

\hangindent=1.5em
\hangafter=1
\noindent Press WH \& Vishniac ET, 1980, ApJ 236, 323.%cited in HS96.

\hangindent=1.5em
\hangafter=1
\noindent Sakharov AD, 1966, Soviet Journal of Experimental and Theoretical Physics 22, 241.%reference from Weinberg 2012 dark energy review.

\hangindent=1.5em
\hangafter=1
\noindent Seljak U \& Zaldarriaga M, 1996, ApJ: 469, 437.%; astro-ph/9603033
%CMBFAST?

\noindent Seo HJ \& Eisenstein DJ, 2003. ApJ 598, 2, 720-740.%http://adsabs.harvard.edu/abs/2003ApJ...598..720S

\hangindent=1.5em
\hangafter=1
\noindent Silk J, 1968, ApJ 151:459-471.

\noindent Sunyaev RA \& Zel'dovich Ya. B, 1970, Ap\&SS 7, 3.

\hangindent=1.5em
\hangafter=1
\noindent Weinberg S, 2002. ApJ 581, 2, 810-816.%http://adsabs.harvard.edu/abs/2002ApJ...581..810W

\hangindent=1.5em
\hangafter=1
\noindent Weinberg DH, Mortonson MJ, Eisenstein DJ, Hirata C, Riess AG \& Rozo
E, 2013, Physics Reports 530, 2, 87-255.%arXiv:1201.2434. 

\hangindent=1.5em
\hangafter=1
\noindent Weinberg S, 2008, Cosmology. University Press: Oxford, UK.

\hangindent=1.5em
\hangafter=1
\noindent Yamamoto K, Sugiyama N \& Sato H, 1997, ApJ 501, 2, 442-460.%http://adsabs.harvard.edu/cgi-bin/bib_query?arXiv:astro-ph/9709247

\section*{Appendix}
\subsection*{Integrals for massless-baryon case (\S\ref{sec:zerobaryons})}
The integral over the growing mode in equation (\ref{eqn:vpsoln}) becomes
\begin{align}
&I_{\rm grow}\left(y,x\right)\equiv \int_{\sqrt{1+\ysc(x)}}^{\sqrt{1+y}} \frac{G(y')\Fg(y')dy'}{W(y')}\nonumber\\
&= \frac{2(5\sqrt{2}-7)}{3\pi} \int_{\sqrt{1+\ysc(x)}}^{\sqrt{1+y}}\frac{\left(3z'^{2}-1\right)\left[ 5+z'\left(4+z'\right)\right]}{\left(z'-1\right)^{2}\left(z' + 1\right)^{3}}dz'.
\end{align}
using the change of variable $z'=\sqrt{1+y'}$, $dy'=2z'dz'$
This can be decomposed into partial fractions and integrated with the substitution $u=z'\pm1$ to yield 
\begin{align}
&I_{\rm grow}\left(y,x\right)=\nonumber\\
& \frac{2(5\sqrt{2}-7)}{3\pi}  \bigg[-\frac{5}{2\left(z'-1\right)}-\frac{1}{2\left(z'+1\right)^{2}}+\frac{1}{1+z'}\nonumber\\
&+\frac{1}{4}\ln\left[\frac{\left(z'-1\right)^{21}}{\left(z'+1\right)^9}\right]\bigg]\bigg|_{\sqrt{1+\ysc(x)}}^{\sqrt{1+y}}.
\label{eqn:I_grow_zero_baryons}
\end{align}

With the same change of variables the integral over the decaying mode in equation (\ref{eqn:vpsoln}) becomes
\begin{align}
&I_{\rm decay}\left(y,x\right)\equiv \int_{\sqrt{1+\ysc(x)}}^{\sqrt{1+y}} \frac{D(y')\Fg(y')dy'}{W(y')}=\frac{3( 7 - 5\sqrt{2})}{\pi} \nonumber\\
&\times\int _{\sqrt{1+\ysc(x)}}^{\sqrt{1+y}}\frac{\left[5+z'\left(4+z'\right)\right]\left[-3z'+\left(3z'^{2}-1\right)\operatorname{arccoth}z'\right]}{\left(z'-1\right)^{2}\left(z'+1\right)^{3}}dz'.
\end{align}
This can be decomposed into partial fractions; the first five terms can be integrated with the same $u$ substition as earlier and the last resulting term has integral
\begin{align}
&I_{\rm last}\left(z'\right)\nonumber\\
&=\frac{3(5\sqrt{2}-7)}{64\pi\left(1+z'\right)}\bigg\{\frac{2}{z'^{2}-1}\bigg(-17-z'\left(61+z'\left(67+15z'\right)\right)\nonumber\\
&+4\operatorname{arccoth}z'\bigg\{11+z'\left(31+z'\left(25+13z'\right)\right)\nonumber\\
&+42(z'-1)\left(1+z'\right)^{2}\operatorname{arccoth}z'\nonumber\\
&+24\left(z'-1\right)\left(z'+1\right)^{2}\ln\left[\frac{2}{1+z'}\right]\bigg\}\bigg)\nonumber\\
&-96\left(1+z'\right)\operatorname{Li}_2\left[\frac{z'-1}{z'+1}\right]\bigg\},
\end{align}
meaning
\begin{align}
&I_{\rm decay}\left(y, x\right)=\nonumber\\
&\bigg[\frac{5\sqrt{2}-7}{4\pi} \left(-\frac{45}{z'-1}+\frac{9}{\left(z'+1\right)^{2}}+\frac{18}{1+z'}+\frac{9}{2}\ln\left[\frac{z'-1}{z'+1}\right]\right)\nonumber\\
&+I_{\rm last}\left(z'\right)\bigg]\bigg|_{\sqrt{1+\ysc(x)}}^{\sqrt{1+y}}.
\label{eqn:I_decay_zero_baryons}
\end{align}

\subsection*{Integrals for massive baryon case (\S\ref{sec:w_baryons})}
Here we evaluate the additional integrals over the baryon forcing $\Fb$ required in \S\ref{sec:analytic}. The baryons' contribution is maximal at decoupling, where it is $\Fb(\yd)/\Fg(\yd)=22.5\%$ for $\fb=0.2$. Using the same change of variables as for the massless baryons, the integral over the growing mode can be decomposed into partial fractions and integrated term by term to yield
\begin{align}
& I_{\rm grow,b}(y, x)\equiv \int_{\sqrt{1+\ysc(x)}}^{\sqrt{1+y}} \frac{G(y')\Fb(y')dy'}{W(y')}=\nonumber\\
&\fb \big[3z'+5\ln[z'-1]-\frac{4}{z'+1}+3\ln[z'+1]\bigg|_{\sqrt{1+\ysc(x)}}^{\sqrt{1+y}}.
\label{eqn:I_grow_baryons_only}
\end{align}

Using the same substitution, the integral over the decaying mode can be decomposed into partial fractions. The first four resulting terms can be integrated directly and the last term has integral
\begin{align}
&I_{\rm last,b}(z')=\frac{9}{4(1+z')}\bigg\{-1-z'(2+3z')\nonumber\\
&+\operatorname{arccoth} z' \bigg[1-z'((3+3z'(7+z'))+10(1+z')\operatorname{arccoth}z'\nonumber\\
&-16(1+z')\ln[\frac{z'+1}{2}]\bigg]-9(1+z')\ln[z'^2-1]\nonumber\\
&-8(1+z')\operatorname{Li}_2\left[\frac{z'-1}{z'+1}\right]\bigg\}.
\end{align}
The integral over the decaying mode is thus
\begin{align}
& I_{\rm decay,b}(y, x)\equiv \int_{\sqrt{1+\ysc(x)}}^{\sqrt{1+y}} \frac{D(y')\Fb(y')dy'}{W(y')}\nonumber\\
&=f_{\rm b}\bigg\{\frac{27}{2}z'+\frac{1}{4}\ln\left[\left(z'-1\right)^{135}\left(z'+1 \right)^{27}\right]\nonumber\\
&-\frac{27}{2}(z'+1)^{-1} +I_{\rm last, b}(z')\bigg\}\bigg|_{\sqrt{1+\ysc(x)}}^{\sqrt{1+y}}.
\label{eqn:I_decay_baryons_only}
\end{align}
%%%%%%%%%%%%%%%%
\end{document}